\documentclass{PoS}
\title{Explosive nucleosynthesis: nuclear physics impact using neutrino-driven wind simulations}

\ShortTitle{Nuclear physics impact on dynamical r-rpocess in neutrino-driven winds}

\author{\speaker{Almudena Arcones}%
  \thanks{Support of
    Swiss National Science Foundation is acknowledged.}\\
  Department of Physics, University of Basel, Klingelbergstra{\ss}e
  82, 4056, Basel, Switzerland\\
  E-mail: \email{a.arcones@unibas.ch}}

\author{Gabriel Mart\'inez-Pinedo\\
  GSI Helmholtzzentrum f\"ur Schwerionenforschung,
  Planckstr. 1, 64291 Darmstadt, Germany\\
  E-mail: \email{g.martinez@gsi.de}}

\abstract{We present nucleosynthesis studies based on hydrodynamical
  simulations of core-collapse supernovae and their subsequent
  neutrino-driven winds.  Although the conditions found in these
  simulations are not suitable for the rapid neutron capture
  (r-process) to produce elements heavier than A$\sim$130, this can be
  solved by artificially increasing the wind entropy. In this way one
  can mimic the general behavior of an ejecta where the r-process
  occurs. We study the impact of the long-time dynamical evolution and
  of the nuclear physics input on the final abundances and show that
  different nuclear mass models lead to significant variations in the
  abundances. These differences can be linked to the behaviour of
  nuclear masses far from stability. In addition, we have analyzed in
  detail the effect of neutron capture and beta-delayed neutron
  emission when matter decays back to stability. In all our studied
  cases, freeze out effects are larger than previously estimated and
  produce substantial changes in the post freeze out abundances.  }

\FullConference{11th Symposium on Nuclei in the Cosmos, NIC XI\\
		July 19-23, 2010\\
		Heidelberg, Germany}

\begin{document}

\section{Introduction}
Half of the elements heavier than iron are produced by rapid neutron
captures in a yet unknown astrophysical scenario.  Galactic chemical
evolution models favor core-collapse supernovae, since they occur
early and frequently enough to account for the abundances observed in
old halo stars and in the solar system
\cite{Ishimaru.etal:2004,Qian.Wasserburg:2007}.  Although the
necessary conditions to produce heavy elements ($A>130$) are
identified \cite{Meyer92} (high entropies, low electron fractions, and
short expansion timescales), these are not found in the most recent
long-time supernova simulations
\cite{arcones.janka.scheck:2007,Fischer.etal:2010}.  When a supernova
explodes, matter surrounding the proto-neutron star is heated by
neutrinos and expands very fast reaching sometimes even supersonic
velocity \cite{Thompson.Burrows.Meyer:2001}.  This neutrino-driven
wind moves through the early supernova ejecta and eventually collides
with it. The interaction of the wind with the slow-moving ejecta
results in a wind termination shock or reverse shock where kinetic
energy is transformed into internal energy. Therefore, the expansion
velocity drops and the temperature (and thus the entropy) increases
after the reverse shock.  The matter near the proto-neutron star
consists mainly of neutrons and protons due to the high temperatures
in this region.  When a mass element expands, its temperature
decreases and neutrons and protons recombine to form alpha
particles. At lower temperatures some of the alpha particles can form
$^{12}$C either by the triple alpha reaction or by the sequence
$\alpha(\alpha n,\gamma){}^9\mathrm{Be}(\alpha,n)^{12}$C. The carbon
nuclei will capture additional alpha particles (alpha-process) until
iron group or even heavier nuclei are
produced~\cite{Woosley.Hoffman:1992}. The amount of these seed nuclei
depends on the entropy and the expansion timescale of the ejecta. Once
the formation of $^{12}$C nuclei freezes out the remaining neutrons
can be captured by the newly formed seed nuclei and the r-process
starts.

\section{Supernova simulations and nucleosynthesis networks}
\label{sec:sn_netw}
For our nucleosynthesis studies \cite{Arcones.Martinez-Pinedo:2010} we
use trajectories, i.e. density and temperature evolutions, from
Ref.~\cite{arcones.janka.scheck:2007}. These are long-time
hydrodynamical simulations that follow the evolution of the explosion
and neutrino-driven winds. Explosions are triggered by neutrinos and
their luminosities are parametrized to obtain typical explosion
energies. The conditions found in the simulations do not allow the
synthesis of heavy r-process elements ($A>130$)
\cite{Arcones.Montes:2010}. Therefore, we need to artificially
increase the neutron-to-seed ratio (by increasing the entropy by a
factor two, which is equivalent to divide the density by a factor of
two) in order to produce the third r-process peak. This allows us to
study the nucleosynthesis of heavy elements in a typical high-entropy
neutrino-driven wind.  At high temperatures, the evolution of the
composition is followed using a full reaction network [10], which
includes nuclei from neutrons and protons to Eu.  Reactions with
neutral and charged particles were taken from the calculations of the
statistical code NON-SMOKER \cite{Rauscher.Thielemann:2000} and
experimental rates were included (NACRE,
\cite{Angulo.Arnould.ea:1999}) when available. The theoretical weak
interaction rates are the same as in
Ref.~\cite{Froehlich.Martinez-Pinedo.ea:2006}.  During the r-process
phase, i.e. after charged-particle freeze-out, we use a fully implicit
network code that includes photodissociation, neutron capture,
beta-decay, and fission. Therefore, it can be used to study the late
evolution when matter decays to stability and the neutron density
becomes very low.

\section{Impact of the nuclear physics input on the dynamical
  r-process}
\label{sec:rprocess}
\begin{figure}[!htp]
  \includegraphics[width=0.47\linewidth]{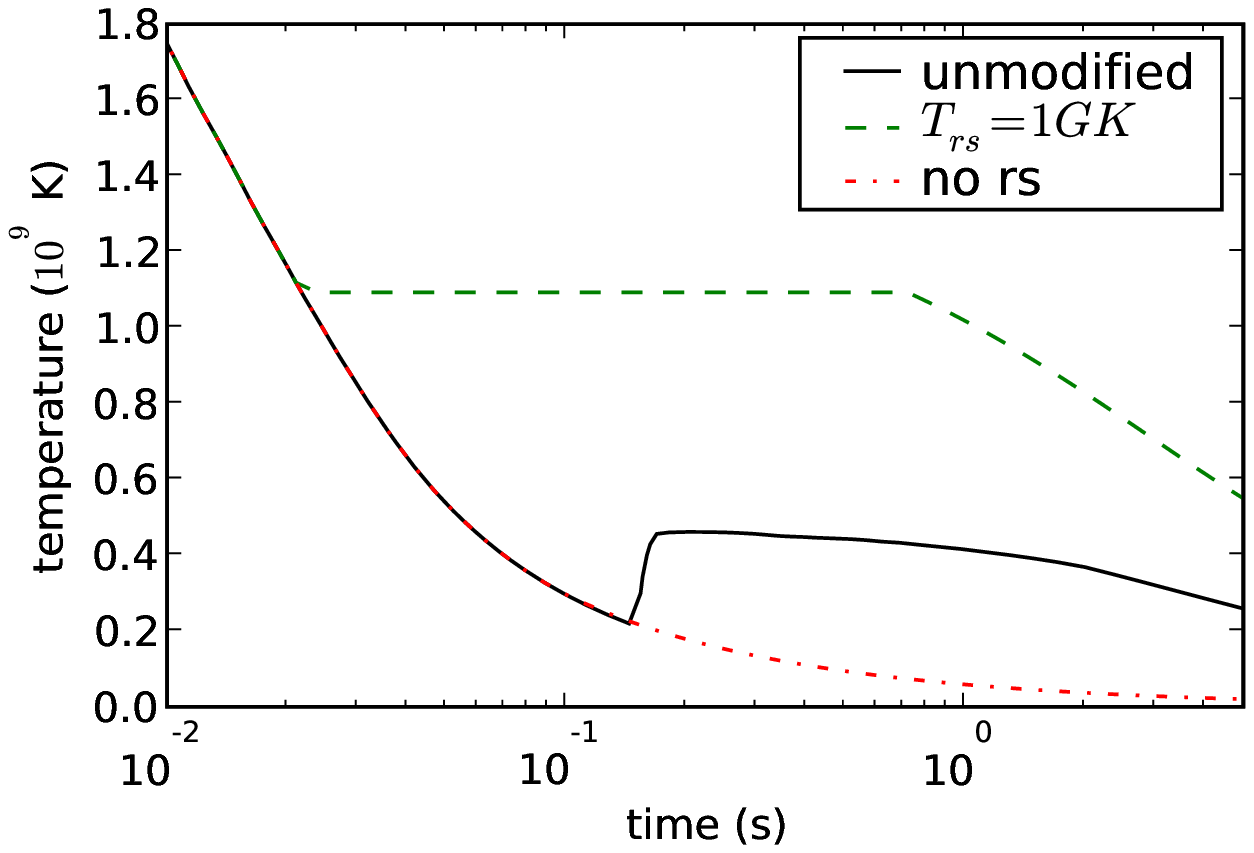}%
  \includegraphics[width=0.47\linewidth]{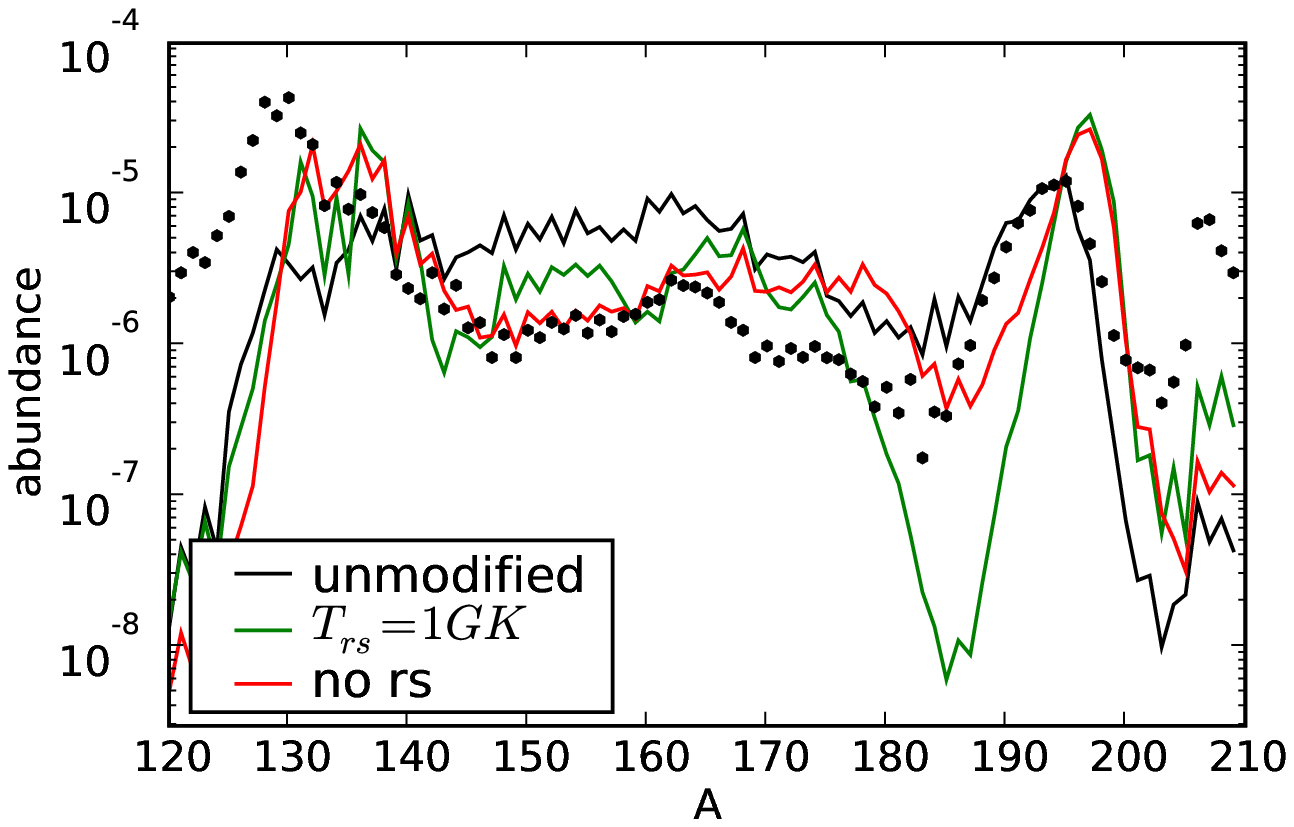}
  \vspace{-0.5cm}
  \caption{{\small Temperature evolution (left panel) of a mass element
    ejected at 8~s after the explosion and variations of the long-time
    evolution. The right panel shows the final abundances (based on
    ETFSI-Q mass model) for the three trajectories and the solar
    abundances by dots.}}
  \label{fig:rs_td}
\end{figure}

We investigate the sensitivity of r-process abundances to the combined
effects of the long-time dynamical evolution and nuclear physics input
and provide a link between the behaviour of nuclear masses far from
stability and features in the final abundances.  
The left panel of Fig.~\ref{fig:rs_td} shows the temperature evolution
during the r-process phase for the three trajectories used in our
calculations: the black line (``unmodified'') corresponds to the
hydrodynamical simulation with the entropy increased and the reverse
shock not changed, in the green dashed line the reverse shock is moved
to a temperature of $T_{\mathrm{rs}}\approx 1$~GK, and for the red
dotted line (``no rs'') the reverse shock was removed. The abundances
resulting from these three evolutions are shown in the right panel of
Fig.~\ref{fig:rs_td}, compared to the solar abundances shown by
dots. Notice that the long time evolution has a big impact on the
position of the peaks and on the troughs. 

One can distinguish two typical evolutions depending on the
temperature: hot ($T=1$GK) and cold ($T<0.5$GK) r-process
\cite{Wanajo:2007}. In the hot r-process the evolution proceeds under
$(n,\gamma )$--$ ( \gamma,n)$ equilibrium which lasts until neutrons
are exhausted, this is similar to the classical r-process
\cite{Kratz.Bitouzet.ea:1993}. For the cold r-process, there is a
competition between neutron capture and beta decay while
photodissociation is negligible. Therefore, the r-process path can
move farther away from stability reaching nuclei with shorter
half-lives which leads to a faster evolution and an earlier freeze
out. Moreover, neutron separation energies have less impact on the
final abundances because they enter only through the neutron capture
cross section. Notice that photodissociation depends exponentially on
the neutron separation energy. The importance of the different nuclear
physics input depends thus on the dynamical evolution, therefore all
our studies are performed for hot and cold r-process
\cite{Arcones.Martinez-Pinedo:2010}.

\begin{figure}[!htp]
  \includegraphics[width=0.5\linewidth,angle=0]{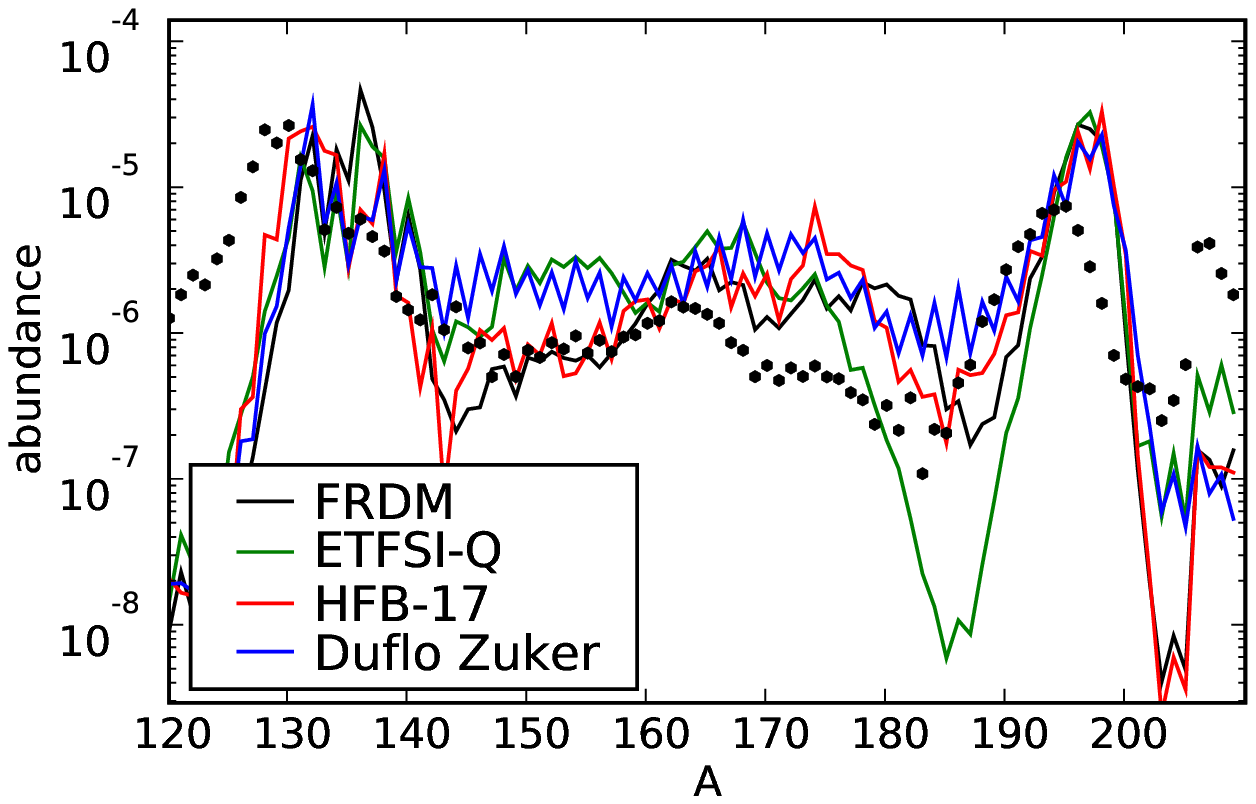}%
  \includegraphics[width=0.5\linewidth,angle=0]{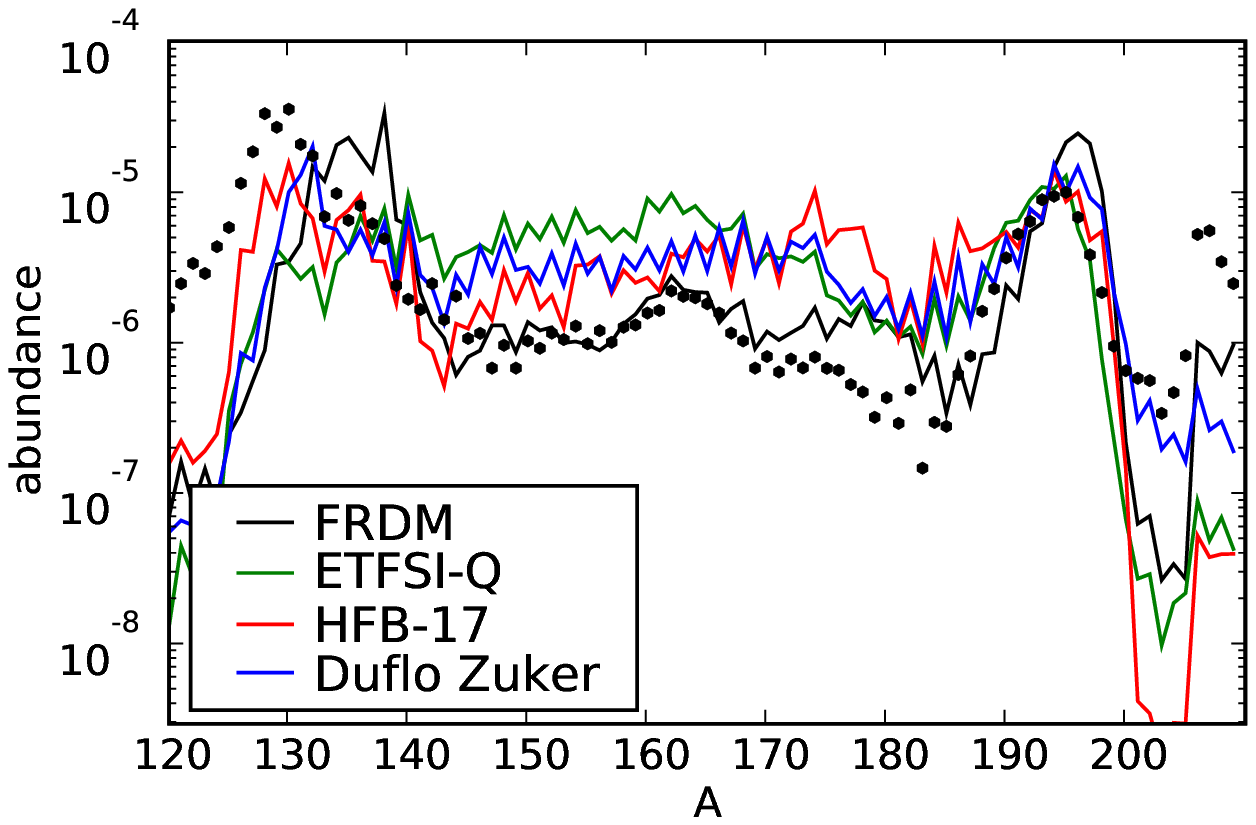}
  \vspace{-0.5cm}
  \caption{{\small Abundances for the mass models indicated in the
      caption and for hot (left) and cold (right) r-process compared
      to solar (dots).}}
  \label{fig:abundmm}
\end{figure}
The sensitivity of the mass model have been investigated by
consistently changing neutron separation energies and neutron capture
rates for the mass models: FRDM~\cite{Moeller.Nix.ea:1995},
ETFSI-Q~\cite{Pearson.Nayak.Goriely:1996},
HFB-17~\cite{Goriely.Chamel.Pearson:2009}, and
Duflo-Zuker~\cite{Duflo.Zuker:1995}.  The final r-process abundances
based on these mass models are shown in Fig.~\ref{fig:abundmm} for hot
and cold r-process conditions. The largest differences in the
abundances are in the region around $A \sim 185$ and can be understood
looking at the behaviour of the two neutron separation energies before
$N=126$ (see Fig.~5 of Ref.~\cite{Arcones.Martinez-Pinedo:2010} which
also provides a deeper analysis of the evolution of the
abundances). Results based on FRDM are affected by the anomalous
behaviour of the neutron separation energy before $N=90$, which
produces the accumulation of matter and thus the formation of peaks
around $A\approx 135$ even for the cold r-process
(Fig.~\ref{fig:abundmm}).

\begin{figure}[!htp]
  \includegraphics[width=0.5\linewidth,angle=0]{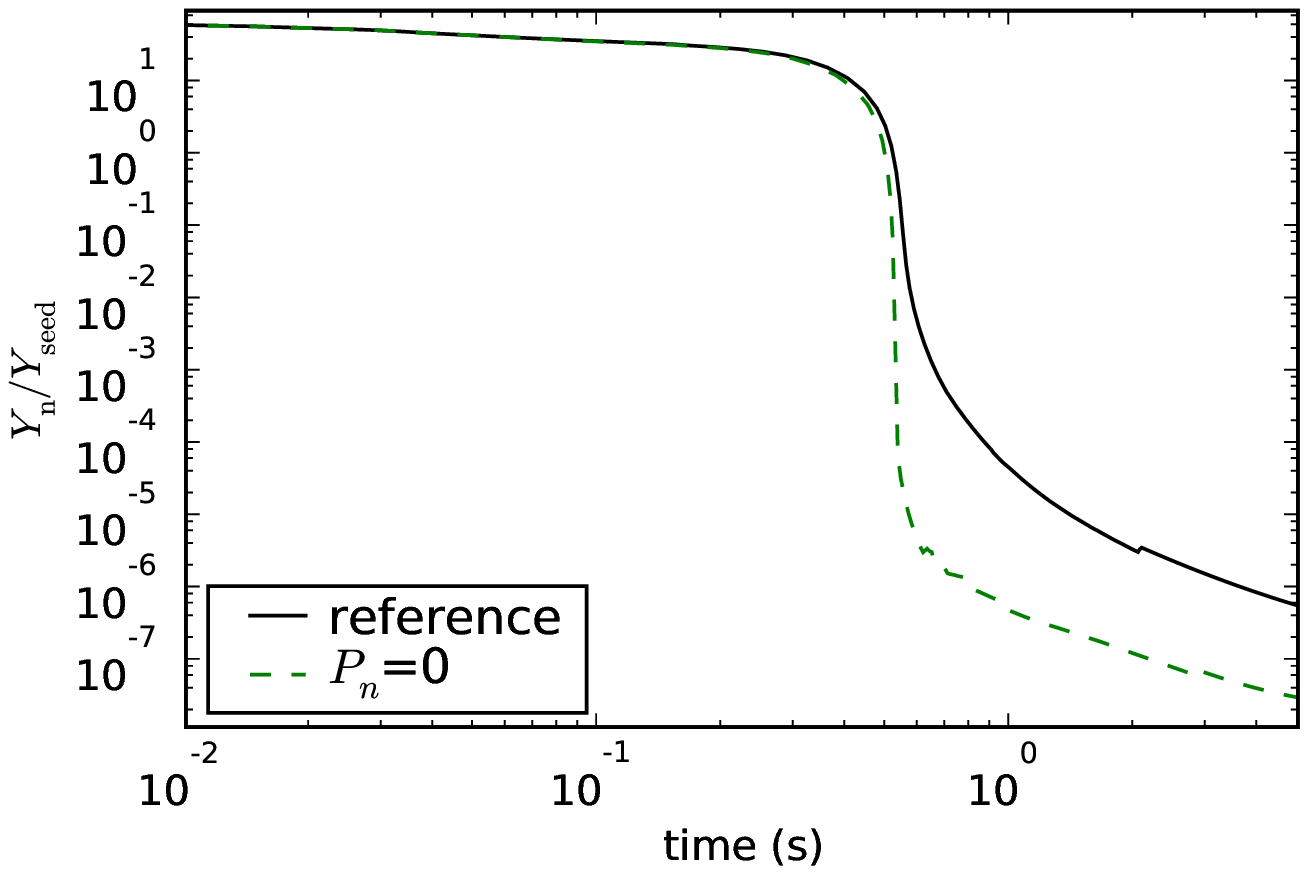}%
  \includegraphics[width=0.5\linewidth,angle=0]{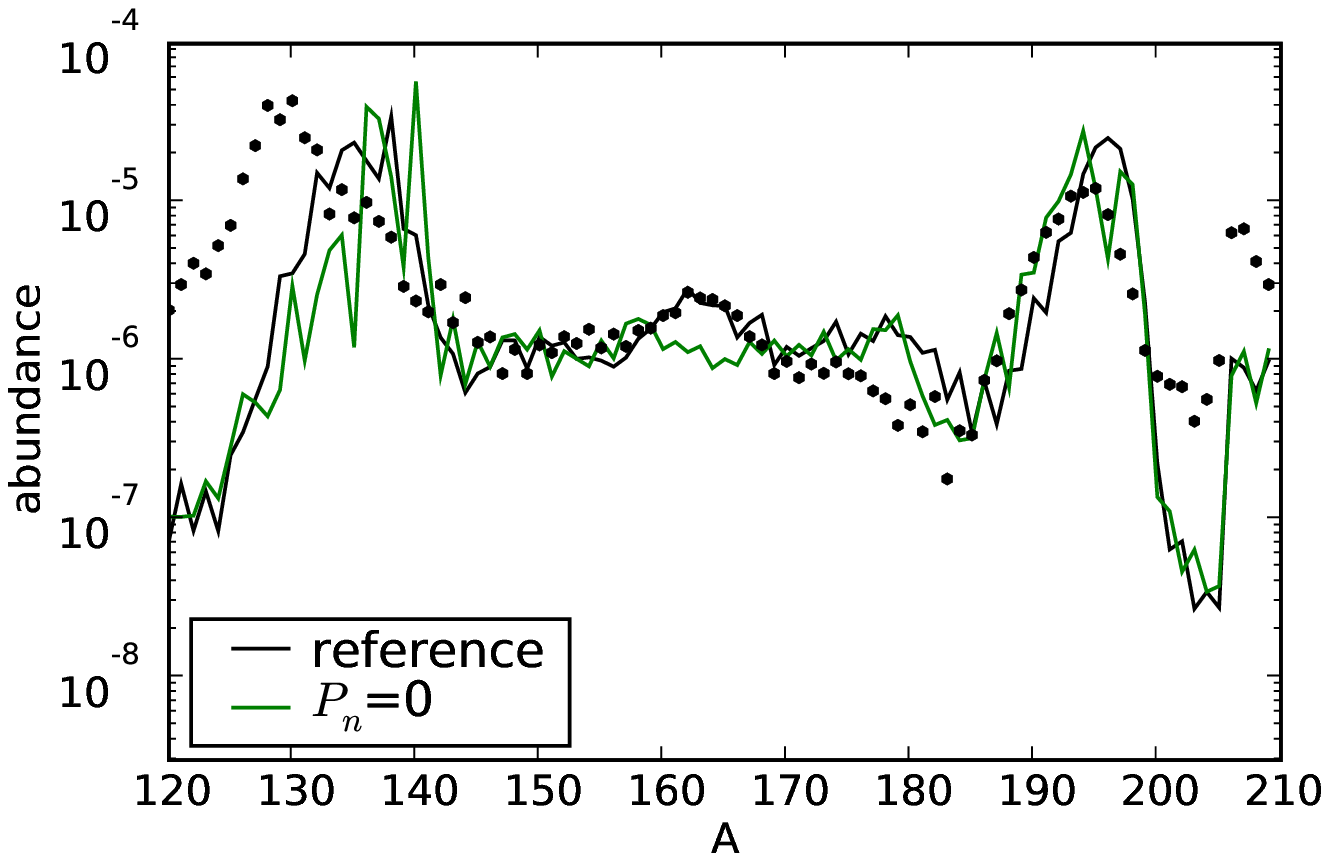}
  \vspace{-0.5cm}
  \caption{{\small Neutron-to-seed ratio and abundances for the cold
      r-process.  The black lines are for the reference case which is
      calculated with the standard nuclear input where neutrons are
      emitted with given probability ($P_n$) after beta decay. The
      green lines are for the case where $P_n=0$, therefore A is
      conserved during beta-decay.}}
  \label{fig:betan}
\end{figure}

In dynamical r-process calculations beta-delayed
neutron emission and neutron captures contribute to the redistribution
of matter, in contrast to the classical r-process calculation (waiting
point approximation) where only the first is considered. The neutron
captures become very important after freeze out (when neutron-to-seed
ratio is around one) as only few neutrons are available and nuclei
compete to capture them.  We find that the rare earth peak is due to
neutron captures when matter moves back to stability, as suggested in
Ref.~\cite{Surman.Engel.ea:1997}. This implies that the freeze-out of
the neutron capture is not instantaneous because neutrons are still
needed to form this feature which is present in the solar r-process
abundances.  Finally, we found that the main contribution of the
beta-delayed neutron emission is the supply of neutrons. In our hot
r-process calculations, there is almost no difference in the abundance
calculated with and without beta-delayed neutron emission because
photodissociation prevents the path to reach the regions far from
stability where the probability of emitting neutrons after beta decay
is higher. In contrast, the suppression of beta-delayed neutron
emission in cold r-process calculations leads to significant changes
in the evolution of the neutron-to-seed ratio. In this case also the
neutron-to-seed ratio (shown in the left panel of
Fig.~\ref{fig:betan}) reaches very small values which produces a minor
shift of the third peak after freeze-out but also inhibits the
formation of the rare earth peak (right panel in
Fig.~\ref{fig:betan}).

\section{Conclusions}
\label{sec:conclusions}
We have explored the impact of the long-time dynamical evolution and
of nuclear masses on the r-process abundances.  We have found that the
relevance of the different nuclear physics inputs depends on the
long-time dynamical evolution \cite{Arcones.Martinez-Pinedo:2010}. If
an $(n,\gamma)$-$(\gamma,n)$ equilibrium is reached (hot r-process),
nuclear masses have a big influence on the final abundances. While for
a cold r-process there is a competition between neutron capture and
beta decay and these two process become relevant.  This rises the
importance of future experiments to measure nuclear masses that will
provide a direct input for network calculations and constraints for
the theoretical mass models.  In both cases, as matter decays to
stability, neutron captures become key to understand the final
abundances and beta-delayed neutron emission becomes important not
only for the redistribution of matter, but also for the supply of
neutrons.  The neutron captures during the decay to stability are
required to explain the rare earth peak. More experimental effort is
necessary to test the validity of the current theoretical cross
sections and more sensitivity studies of the impact of the neutron
capture rates on the final abundances will give rise to new insights.


\providecommand{\newblock}{}

\end{document}